\documentclass[rnote,traditabstract]{aa} 

\usepackage{amssymb}
\usepackage{amsmath}

\usepackage{graphicx}
\usepackage{graphicx}
\usepackage{epsfig}
\usepackage{graphics}
\title{Transition disks: 4 candidates for ongoing giant planet formation in Ophiuchus}
\author{Mariana Orellana\inst{1,2},
Lucas. A. Cieza\inst{3}, 
Matthias R. Schreiber\inst{2},
Bruno Mer\'{\i}n\inst{4},
Joanna M. Brown\inst{5},\\
Leonardo J. Pellizza\inst{6,1},
Gisela A. Romero\inst{2}
}
\institute{Consejo Nacional de Investigaciones Cient\'{\i}ficas y Te\'cnicas (CONICET), Argentina.
\and Departamento de F\'{\i}sica y Astronom\'{\i}a, Universidad de Valpara\'{\i}so, Av. Gran Breta\~na 1111, Valpara\'{\i}so, Chile.
\and Institute for Astronomy, University of Hawaii at Manoa,  Honolulu, HI 96822, \emph{Sagan} Fellow.
\and Herschel Science Centre, European Space Astronomy Centre (ESA), P.O. Box 78, 28691 Villanueva de la Ca\~nada (Madrid), Spain.
\and Harvard-Smithsonian Center for Astrophysics, 60 Garden St. MS 78, Cambridge, MA 02138.
\and Instituto de Astronom\'{\i}a y F\'{\i}sica del Espacio, C.C. 67, Suc. 28, (1428) Buenos Aires, Argentina.}

\offprints{Mariana Orellana {\em morellana@fcaglp.unlp.edu.ar}}
\titlerunning{Four planet forming candidates in Ophiuchus}

\begin{document}
\abstract{
A large set of \emph{Spitzer}-selected transitional disks in the Ophiuchus molecular cloud was recently examined by Cieza et al (2010), and 4 of the targets were identified as (giant) planet-forming candidates based on the morphology of their Spectral Energy Distributions (SEDs), the apparent lack of stellar companions, and evidence for accretion. Here we characterize the structures of these disks modeling their optical, infrared and (sub)millimeter SEDs.  We use the  Monte Carlo radiative transfer package RADMC to construct a parametric model of the dust distribution in a flared disk with an inner cavity and calculate the temperature structure consistent with the density profile, in thermal equilibrium with the irradiating star. For each object, we conducted a Bayesian exploration of the parameter space generating Monte Carlo Markov Chains (MCMC) that allow the identification of the best fit parameters and to constrain their range of statistical confidence. Our calculations point to the presence of evacuated  cavities with radii $\sim2-8$ AU, consistent with having been carved by embedded giant planets. 
We found parameter values consistent with those previously given in the literature, indicating a mild degree of grain growth and dust settling, which deserves to be investigated with further modeling and follow up observations. Resolved images with (sub)millimeter interferometers are required to break some of the degeneracies of the models and better constrain the physical properties of these fascinating disks. 
}
\authorrunning{M. Orellana et al.}
\date{Accepted 22-11-2011}
\maketitle
%
\section{Introduction}\label{intro}
The infrared spectral energy distributions (SEDs) of circumstellar transitional disks reveal the presence of an optically thin inner region and an optically thick outer disk. 
Several mechanisms relevant to the overall evolution of circumstellar disks, and in particular to the short-lived phase when they dissipate, have been proposed to explain the so-called opacity holes of  transition disks:  giant planet formation, grain growth, photoevaporation, and tidal truncation in close binaries. See Williams \& Cieza, 2011 for a  recent review.
The processes responsible for the inner holes of transition disks can tentatively be distinguished when disk masses, accretion rates, and multiplicity information are available (Najita et al. 2007; Cieza  2008). 
Following this approach, Cieza et al. (2010, hereafter Paper~I) presented the initial results of an ongoing project  to characterize a large set of transition disk candidates in nearby star-forming regions.\\ 

Probing the structure of disks that are suspected to be forming planets is the most promising approach to understand the conditions in which planets are formed. 
The best indication for ongoing planet formation in disks is the detection of tidal gaps (e.g. Pi\'etu et al. 2006) corresponding to a ring with significant decrease in the surface density (or the whole inner disk if it was depleted by accretion, e.g. Varnei\'ere et al 2006). 
A spectacular confirmation has arrived with the recent detection of the first potential substellar object within the gap of the transitional disk T Chamaleontis (Hu\'elamo et al. 2011).
Inner holes and gaps have already been observed at (sub)millimeter wavelengths in a handful of objects bright enough for resolving disk structure
 (Pi\'etu et al. 2006; Hughes et al. 2007, 2009; Brown et al. 2008, 2009; Andrews et al. 2009, 2010a, 2011, Isella et al 2010a, 2010b).\\

Comprehensive studies of similar transition disks are necessary to increase the empirical constraints on their structures and investigate their diversity. This is the primary motivation of this work, where we apply a parametric description to the 4  planet-forming disks candidates in Ophiuchus presented in Paper~I and derive the best fitting physical characteristics for their known SEDs.


\section{The targets}
In Paper~I, the observed SEDs were characterized by two parameters, as introduced by Cieza et al. (2007): the longest wavelength at which the observed flux is dominated by the stellar photosphere, $\lambda_{\rm turn-off}$, and the slope of the IR excess,  $\alpha_{\rm excess}$, computed between $\lambda_{\rm turn-off}$ and 24 $\mu$m. The former parameter correlates with the size of the inner hole, and the latter with the sharpness of the edge of the hole, i.e., a large increase in the dust density over a small range in radii is indicated by a positive $\alpha_{\rm excess}$. Accreting disks with sharp inner holes that seem to lack stellar companions (e.g., from adaptive optics imaging) stand as the most promising candidates for ongoing (giant) planet formation. 

There were 4 cases fulfilling these criteria in Paper I  from a sample of 26 of \emph{Spitzer}-selected transitional disks in the Ophiuchus molecular cloud ($d\sim 125$ pc, Loinard et al. 2008). 
Their basic properties are listed in Table \ref{properties}.  The mass accretion rates for our targets, inferred from the H$\alpha$ emission line widths (Natta et al. 2004), range from 1.3$\times$10$^{-10}$  to 5$\times$10$^{-8}$ M$_\odot/$yr.

\begin{table*}[!t]
\caption{Target Properties}
\begin{center}
\begin{tabular}{lcccccccccccc}
\hline
\noalign{\smallskip}
{Name$^a$} & {\emph{Spitzer} ID} &  {SpT} &  {$T_{\rm eff}$} &  {$L_\star$} &  {$R_\star$}&  {$M_\star^b$} & {Age} & {A$_V$} &  {$\log\dot{M}_{\rm acc}^c$} &  {$\lambda_{\rm turn-off}$} &  {$\alpha_{\rm excess}$}&  {$M_{\rm disk}^d$}\\
 {} & {SSTc2d\_} & {} &  {(K)} &  {($L_{\odot}$)} &  {($R_{\odot}$)}&  {($M_{\odot}$)} &  {(Myr)} &  {(mag)} &  {($M_{\odot}$/yr)} &  {($\mu$m)} & {} & {($M_{\rm JUP}$)}\\
\noalign{\smallskip}\hline \noalign{\smallskip}
Tran 11 & J162506.9-235050 & M3 &  3470  &  0.24  & 1.25  &  0.3  & 2.1 & 3.8 & -8.8 &  8.0 &  0.65 & $<$1.5\\  
Tran 21 & J162854.1-244744 & M2 &  3580  &  0.51  & 1.74  &  0.4  & 1.4 & 5.4 & -9.3 &  8.0 &  0.69 & 1.3 \\    
Tran 31 & J163205.5-250236 & M2 &  3580  &  0.19  & 1.08  &  0.4  & 4.1 & 5.0 & -7.3 &  8.0 &  0.30 & $<$1.3\\  
Tran 32 & J163355.6-244205 & K7 &  4000  &  0.78  & 1.70  &  0.7  & 2.0 & 5.0 & -9.9 &  8.0 &  0.72 & 11.1\\    
\noalign{\smallskip}
\hline\noalign{\smallskip}
\multicolumn{13}{c}{\parbox[b]{\textwidth}{$^a$ Alternative names: Tran 21 is WSB 63, Tran 31 is WSB 75, and Tran 32 is RXJ1633.9-2242. 
$^b$ Stellar parameters from pre-main sequence evolutionary tracks by Siess et al (2000).
$^c$ Based on the velocity dispersion of the H$\alpha$ line from Paper~I.
$^d$ Rough estimates of the  disk masses based on a single (sub)millimeter  flux or upper limits from Paper~I.}}
\end{tabular}
\end{center}
\label{properties}
\end{table*}
For all of them we have at hand the photometric fluxes in the $R$-band from the USNO-B1 catalog, 
the $J$, $H$, and $K_S$ bands, from  the Two Micron All Sky Survey (2MASS), as well as fluxes at 3.6, 4.5, 5.8, 8.0, 24, and 70 $\mu$m obtained by \emph{Spitzer}\footnote{two of the 70 $\mu$m measurements given in Paper~I are upper limits.}. We have assumed errors that are within observational standards ($\la 20 \%$) .  
In addition, all of our targets have been observed with the Submillimeter Array (SMA). For two of them (Tran 21 and 32) we obtained the fluxes at 1.3~mm,  while for Tran 11 and 31 we could only derive upper limits (see Paper I for details on SED data). For Tran 32, $F_{850\,\mu{\rm m}}$ was given in Nutter et al. (2006). 
One of our targets, Tran 11, was recently included in the study of {\em cold} disks (disks with large inner dust holes) performed by Mer\'{\i}n et al (2010), i.e. their
source \# 24. It was modeled including a positive detection from MIPS photometry: $F_{70 \mu{\rm m}}= 537\pm78.8$ mJy, and the measured \emph{Spitzer}-IRS spectrum, but no constraints of the flux at (sub)-millimeter wavelengths.
For the sake of completeness,  we have redone the fit to the SED of  this source with our set of free parameters and including the 1.3 mm 
flux upper limit. 

\section{Radiation transfer models}

The modeling that we have applied is similar to those performed by Andrews et al. (2009, 2010a) and Brown et al. (2009), who have confirmed their physical estimates from SED modeling through direct imaging.
A 2--D structure model for flared disks is combined with the Monte Carlo continuum radiative transfer package RADMC v3.1 (Dullemond \& Dominik, 2004), modified to include a density reduction as an inner cavity.  The code computes a temperature structure consistent with the given density profile, and in equilibrium with the irradiation by the central star. The disk is presumed to be passive, an assumption that  is supported by the low disk to stellar luminosity ratios of our sample, $\la$0.005 according to estimates in Paper~I.\\
We consider a surface density profile characterized by a power-law, $\Sigma \propto R^{-\gamma}$, 
with an exponential taper at larger radii ($\propto e^{-(R/R_c)^{2-\gamma}}$), 
where $R_c$ is the characteristic radius. This is physically motivated by the success of similarity solutions of viscous disks to reproduce the observed gradual density decay at large radii (Hughes et al. 2008).
$\Sigma$ is normalized to obtain the total mass of the disk, $M_d$, when integrated. 
The radial index was fixed to be $\gamma=1$ which is a typical value within the range $\gamma = 0.4 - 1.1$ established by Andrews et al (2010a). Our option could be questioned in the light of results by Isella et al. (2009) who have independently inferred slopes from steep to quite shallow in their sample of spatially resolved disks.\\
Resolved images are therefore mandatory to obtain more accurate estimates of $\gamma$ for our particular targets. We set the characteristic radius $R_c= 100$ AU. However, there is no spatially resolved information in the SEDs alone, and the data can be reproduced equally well with a wide range of outer disk values (Andrews et al. 2010b). The value we choose is representative of the disks with resolved interferometric visibilities, that are $R_c=14-198$ AU (Andrews et al. 2009, 2010b) in Ophiuchus, and $R_c\simeq 30-230$ AU for Taurus-Auriga (Isella, Carpenter \& Sargent, 2009). Note that larger outer radii (100 - 1100 AU) have been obtained with different fitting techniques (sharply truncated power law fits to CO observations) and are not directly comparable with the $R_c$ values (see Williams \& Cieza 2011). Aside from the extreme case of a nearly edge-on viewing angle, the disk inclination cannot be determined from unresolved observations. Scattered light images have proben useful in this sense (Pinte et al. 2007). We have set an intermediate representative inclination $i=30^\circ$ in our modeling.\\
There is strong observational evidence that circumstellar dust can present some degree of settling to the midplane (Furlan et al. 2006, McClure et al. 2010), which is in agreement with theoretical predictions (e.g., Dominik et al. 2007 and references). Indeed, the growth of the dust grains, which proceeds in a complex fashion, can be accelerated by dust settling, and it is expected that the larger particles are aggregated close to the equatorial plane.
The dust can therefore be distributed differently than the gas, which is expected to remain in hydrostatic equilibrium (Chiang \& Goldreich, 1997). This is described by a vertical Gaussian density profile with scale-height $H$ and radial index $\psi$, i.e. $H\propto R^{1+\psi}$, anchored at $R=100$ AU by $H_{100}$. Vertically extended disks (with larger $H$) are heated more efficiently and thus reemit more radiation at mid- and far-IR wavelengths. Here we have fixed $\psi=0.2$ in order to reduce the number of free parameters of the models.\\
Following Andrews et al. (2009), we consider a radius, $R_{\rm cav}$, such that the modified surface density is $\Sigma^{\prime} = 
\delta_{\rm cav} \Sigma$ when $R \le R_{\rm cav}$, and $\delta_{\rm cav} < 1$ artificially reduces the density inside the cavity, in order to mimic the deficit in the disk emission. The whole disk extends down to the distance, $R_{in}$, where the temperature is enough to sublimate the dust grains ($\sim 1500$ K).  Our 4 targets present a small or negligible NIR excess above the photospheric value (note the $\lambda_{\rm 
turn-off}$ values in Table \ref{properties}), suggesting cavities that are  largely evacuated of dust, therefore we have varied $\log(\delta_{\rm cav})$ as a sensitive parameter to be established by the data.\\
We have adopted the opacity spectrum given by Andrews et al. (2009) who considered the silicate and graphite abundances determined for the ISM, and updated optical properties. The grain size distribution, $n(a) \propto a^{-3.5}$, extends in diameters from $a_{{\rm min}} = 0.005$\,$\mu$m up to $a_{\rm max}$=1 mm, and the opacities were calculated with a simple code for Mie scattering.

The stellar properties given in Table \ref{properties} are in each case fixed inputs for the models, and Kurucz spectra are used as models of the central stars. As the disk SED is not influenced by the $\lambda\la 3\, \mu$m fluxes, they were not included in the final fit, avoiding a systematic error term (shift in the $\chi^2$). For each disk, we have explored the parameter space, $\{ M_{\rm disk}, \, H_{100}, \, \, R_{\rm cav}, \, \log \delta_{\rm cav}\}$, by generation of Monte Carlo Markov Chains, which is a parameter space exploration technique designed to provide the best fitting values and their uncertainties. This technique attains a better performance than classical data fitting methods (e.g., Press et al. 1992; Gregory 2005) for problems with large numbers of parameters, which is our case. 
In the fitting procedure we have followed Ford (2005) and Gregory (2005). The marginal probability distribution for each parameter where derived from the MCMC chains. The 90\% percentile of such distributions\footnote{Note that there is no guarantee that the marginal distribution is Gaussian so the dispersion of the sample can only be a rough indicator of the probable error as in Isella et al (2009).} simmetrically centered in the mode when possible (or using the larger side if not) was used to define the error estimates within a conservative approach.

\section{Results}

\begin{figure}[t]
\begin{center}
\includegraphics[width=0.75\linewidth]{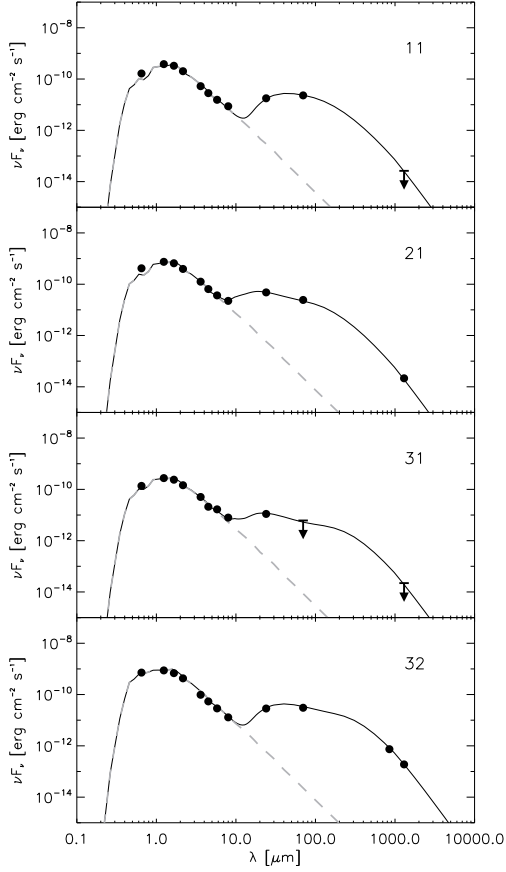}
\caption{Best fit SEDs (solid lines) obtained using RADMC, in the context of a flared disk model with an inner cavity. The dashed line shows the input stellar model atmosphere. The filled circles are extinction corrected values, while the arrows are upper limits.}
\label{own}
\end{center}
\end{figure}

The estimated parameters that best reproduce the data for our planet forming disk candidates are listed in Table~\ref{resul}, while 
the corresponding SEDs are shown in Figure \ref{own}. These fits have reduced $\bar{\chi}^2 = 0.9 - 10$, i.e. the $\chi^2$  considering only the observational data point errors, and divided by the number of degrees of freedom.
The disk masses obtained are within a factor $\sim$2 than the rough estimates from Paper~I, which were based on a single (sub)millimeter flux. 
We find $H_{100}$  $\lesssim$ 2 -- 6 AU in all four cases, with estimated uncertainties i.e. $\la 1.5$ AU. The comparison between the dust scale heights to the scale heights of the gas, which is in hydrostatic equilibrium, give a settling ratio 0.13--0.25.
Specifically for Tran 32 we get a rather flat geometry, with the smallest $H_{100}$ in our sample. This disk presents also the largest cavity and is bright enough to be detected by SMA extended configuration. A simultaneous SED + image fit will be presented in a future paper.

To test our results against an alternative model,  we have used the precomputed grid of SEDs by Robitaille et al. (2007). A cavity is also inferred in all 4 cases, with $R_{\rm cav}\sim 4-38$ AU, as well as some dust settling leading to flat geometries ($H_{100}\la 4$ AU). However, the on-line fitting tool fails to accurately reproduce the SED for 2 of our targets, Tran 11 and 32. A discussion on the preferential use of RADMC to model transitional disks and comments on alternative observational characterization of them can be found in Mer\'{\i}n et al (2010).

\begin{table}[h]
\caption{Model parameters.}
\begin{center}
\begin{tabular}{ccccc}
\hline
\noalign{\smallskip}
\hline
\noalign{\smallskip}
 {\#} &  {$M_{disk} ^{a}$} &  {$H_{100}^{b}$}& {$R_{\rm cav}$} &{$\log \delta_{\rm cav}$}\\
 {}&  {($M_{JUP})$} & {(AU)}&  {(AU)}&  {(dex)}\\
\noalign{\smallskip}\hline
\noalign{\smallskip}
11& $<\!$ 2.0 $\pm$ 0.7  &  $<\!$ 4.2 $\pm$ 1.3 & 4.8 $\pm$ 2.5  & $\la$ -6.2 \\ 
21&  0.6 $\pm$ 0.2  & 3.1 $\pm$ 0.8 & 1.9 $\pm$ 0.3  & -4.9 $\pm$ 0.8\\
31& $<\!$ 1.7 $\pm$ 0.1  & $<\!$ 1.9 $\pm$ 0.1 & 1.5 $\pm$ 0.4  & -5.2 $\pm$ 0.5\\
32& 17.0 $\pm$ 5.3  & 2.0 $\pm$ 0.5 & 7.9 $\pm$ 2.3  & $\la$ -6.3 \\
\noalign{\smallskip}\hline\noalign{\smallskip}
\multicolumn{5}{l} {\parbox[b]{\columnwidth}{$^a$ M$_{\rm disk}$ are upper limits for Tran 11 and 31 as
their  1.3 mm fluxes are also upper limits. $^b$ $H_{100}$ is an upper limit for Tran  31 as
its 70 $\mu$m fluxes is also an upper limit.}}
\end{tabular}
\end{center}
\label{resul}
\end{table}

\section{Discussion}
Extensive work in non-linear model fitting have been applied in many areas, and recently to disk modeling by Isella et al. (2009). They have used the two layer approximation of Chiang \& Goldreich (1997), whereas we compute the thermal structures with significant detail using the RADMC code. In this study we have followed the literature by Andrews et al. (same parametric model and opacities) who have performed refined grid searches over 8 parameters. Their acute comments on the technical obstacles (Andrews et al. 2011) and non-uniqueness of the fit, also apply for our approach.

We have estimated some of the physical parameters describing the dust structures of 4 transition disk systems that are excellent candidates for ongoing giant planet formation.  
The cavities sizes inferred for our targets, $R_{\rm cav}=2-8$ AU, are in agreement with the distribution of semi-major axis of exoplanets, which has a bi-modal behavior, peaking at $\sim 0.05$ and $2$ AU and extending up to $\sim 10$ AU (exoplanets database, {\sf http://exoplanets.org}). 
The mass of the disks and the actual size of their inner holes depend on the assumed opacities. 
The small scale height of the dust in all our targets suggest this could be a characteristic property of planet forming disks. This is an intriguing results that deserved to be investigated with further modeling  and follow up observations. 

An effort was done here ($\sim 10^4$ model runs for disk) to provide the set of best fitting parameters and their confidence regions through the application of Bayesian methods, but the real weight of the results could be yet unclear given the degeneracy of the model into some of the parameters we have fixed. 
An illustration is the case of Tran 11, for which a cavity radius $R_{\rm cav}=3\pm 2$ AU, and a settling parameter that translates into $\psi=0.07$ have been found by Mer\'{\i}n et al. (2010) when setting $R_c=200$ AU. We used $R_c=100$ AU, $\psi=0.2$ and obtained $R_{\rm cav}=4.8\pm 2.5$ AU. 

The disk masses obtained here cover a rather wide range of masses, from $\sim 0.6 M_{\rm JUP}$ to $\sim 17 M_{\rm JUP}$ that extends to lower values the mass range obtained by Andrews et al., i.e. $5-42 M_{\rm JUP}$ for systems that have been modeled in the same way, i.e. with same parameterizations and opacity tables. However, Andrews et al. (2009, 2010a) fitted their systems with larger values of the depletion factor, $\delta_{\rm cav}\sim 10^{-4}-10^{-2}$, and obtain larger cavities, $R_{\rm cav}=20-40$\,AU. We note that in Andrews et al. (2011) where the model include a more complex surface density profile (i.e. with an inner disk inside the cavity) the range of masses estimated for 12 disks with resolved images is $\sim 8 - 128 M_{\rm JUP}$ but comparisons could be misleading in this case.
Some differences are probably the result of a selection effect. While Andrews et al. selected their targets based on large (sub)millimeter fluxes, our selection criterion is based on the slope of IR SEDs, and the presence of accretion.

At a southern declination of $\sim$25 deg and a distance of $\sim$125 pc, our targets are excellent targets for follow up studies with \emph{Herschel} and the Atacama Large Millimeter Array (ALMA) to investigate their properties in more detail, and in particular, the small holed transition disks will require its better spatial resolution.
\emph{Herschel}  far-IR photometry would bridge the gap between the mid-IR and the (sub)mm wavelengths accessible from the ground and help to better constrain the scale heights and the flaring angles of the disks. \emph{Herschel}  spectroscopy of fine structure lines, such as  the 63.2 $\mu$m [O~I] line, could help to probe their gas content and hence the gas to dust mass ratio. 
Similarly, resolved images with ALMA will break some of the degeneracies of the models and will allow to better understand  the physical properties of these fascinating disks, thereby helping to elucidate the conditions in which planets are formed. 


\begin{acknowledgements}
We acknowledge very useful discussions with C. Dullemond and S. Andrews. We thank Herv\'e Bouy for explanations. M.O. was supported by ALMA-CONICYT (31070021) and ANPCyT PICT 2007 00848/Prestamo BID. MRS acknowledges support from Millennium Science Initiative, Chilean ministry of Economy: Nucleus P10-022-F.
\end{acknowledgements}

{}

\begin{thebibliography}{}

\bibitem{} Andrews, S.~M., \& Williams, J.~P. 2005, \apj, 631, 1134 
\bibitem{} Andrews, S.~M., \& Williams, J.~P. 2007, \apj, 671, 1800
\bibitem{} Andrews, S. M., Wilner, D. J., Hughes, A. M., Qi, C., Dullemond, C. P. 2009, \apj, 700, 1502
\bibitem{} Andrews, S. M., et al, 2010a, \apj, 723, 1241 
\bibitem{} Andrews, S. M., et al, 2010b, \apj, 710, 462
\bibitem{} Andrews, S. M., et al, 2011, \apj, 732, 42
\bibitem{} Brown, J. M., Blake, G. A., et al. 2008, ApJ, 675, L109
\bibitem{} Brown, J. M., Blake, G. A., et al. 2009, ApJ, 704, 496
\bibitem{} Chiang, E.~I. \& Goldreich, P., 1997, \apj, 490, 368
\bibitem[Cieza et al.(2007)]{2007ApJ...667..308C} Cieza, L., et al. 2007, \apj, 667, 308 
\bibitem{} Cieza, L.~A. 2008, ASPC, New Horizons in Astronomy, 393, 35 
\bibitem{} Cieza, L.~A. et al, 2010, \apj,  712, 925 (Paper I)
\bibitem{} Dominik, C. et al, 2007, Protostars and Planets V, University of Arizona Press, p.783-800
\bibitem{} Dullemond, C.~P., \& Dominik, C., 2004, A\&A, 417, 159
\bibitem{} Ford, E. B., 2005, ApJ, 129, 1706
\bibitem[Furlan et al.(2006)]{2006ApJS..165..568F} Furlan, E., et al.\ 2006, \apjs, 165, 568 
\bibitem{} Gregory, P., 2005, Bayesian Logical Data Analysis for the Physical Sciences, Cambridge University Press
\bibitem{} Hu\'elamo, N. et al. 2011, A\&A, 528, L7
\bibitem{} Hughes, A. M., Wilner, D. J., Calvet, N., et al. 2007, \apj, 664, 536
\bibitem{} Hughes, A. M., Andrews, S. M., et al. 2009, \apj, 698, 131
\bibitem{} Isella, A.; Carpenter, J. M.; Sargent, A. I. 2009, \apj, 701, 260
\bibitem{} Loinard, L., et al. 2008, \apj, 675, L29
\bibitem[McClure et al.(2010)]{2010ApJS..188...75M} McClure, M.~K., et al.\ 2010, \apjs, 188, 75
\bibitem{} Mer\'{\i}n, B. et al, 2010, \apj, 718, 1200
\bibitem{} Najita, J.~R., Strom, S.~E., \& Muzerolle, J. 2007, \mnras, 378, 369
\bibitem{} Natta, A. et al, 2004, A\&A, 424, 603
\bibitem[Nutter et al.(2006)]{2006MNRAS.368.1833N} Nutter, D., Ward-Thompson, D., \& Andr{\'e}, P.\ 2006, \mnras, 368, 1833 
\bibitem{} Pi\'etu, V., et al 2006, A\&A, 460, L43
\bibitem{} Pinte, C., et al 2007, \apj, 673, L63
\bibitem{} Press, W. H. et al. C Numerical recipes, Cambridge University Press 
\bibitem[Siess et al.(2000)]{2000A&A...358..593S} 
Siess, L., Dufour, E., \& Forestini, M.\ 2000, \aap, 358, 593 
\bibitem{} Robitaille, T. P., Whitney, B. A., Indebetouw, R., Wood, K. 2007, ApJS, 169, 328
\bibitem{} Varni\'ere, P., Blackman, E. G., Frank, A. \& Quillen, A. C., 2006, \apj, 640, 1110
\bibitem{} Williams, J. P, \& Cieza, L. A. 2011, A\&ARA, 2011, 49, 67
\end{thebibliography}
\end{document}